\documentclass[aps,10pt,prd,notitlepage,tightenlines,nofootinbib,superscriptaddress]{revtex4-2}

\usepackage{amsmath}
\usepackage{amssymb,amsthm}
\usepackage{mathtools}
\usepackage{mathrsfs}
\usepackage{relsize}
\usepackage{bm}
\usepackage{ulem}

\usepackage{epsfig,graphics,graphicx,color,xcolor}
\usepackage{soul} 
\usepackage{cancel} 
\usepackage{slashed}	
\usepackage{subfigure}
\usepackage{array}
\usepackage{ragged2e}
\usepackage{lineno}
\usepackage{natbib}
\usepackage{hyperref}
\hypersetup{colorlinks=true,linkcolor=darkgreen,anchorcolor=darkgreen,citecolor=darkgreen, filecolor=brown,urlcolor=darkgreen,bookmarksnumbered=true,
pdfview=FitB
}
\graphicspath{ {images_pert/} }
\usepackage{enumitem}

\colorlet{darkgreen}{green!50!black}
\colorlet{brightyellow}{yellow!75!red}
\colorlet{orange}{red!50!yellow}
\colorlet{darkgray}{gray!50!black}

\def\dd{{\mathrm{d}}}

\makeatletter
\newcommand*{\transpose}{%
  {\mathpalette\@transpose{}}%
}
\newcommand*{\@transpose}[2]{%
  \raisebox{\depth}{$\m@th#1\intercal$}%
}
\makeatother 



\begin{document}

\title{Energy momentum tensor on and off the light cone: exposition with scalar Yukawa theory}

\author{Xianghui~Cao}
\affiliation{Department of Modern Physics, University of Science and Technology of China, Hefei 230026, China}

\author{Siqi~Xu}
\affiliation{Institute of Modern Physics, Chinese Academy of Sciences, Lanzhou 730000, China}
\affiliation{School of Nuclear Science and Technology, University of Chinese Academy of Sciences, Beijing 100049, China}

\author{Yang~Li}
\affiliation{Department of Modern Physics, University of Science and Technology of China, Hefei 230026, China}
\affiliation{Department of Physics and Astronomy, Iowa State University, Ames, Iowa 50011}

\author{Guangyao~Chen}
\affiliation{Department of Physics, Jacksonville University, Jacksonville, FL, 32211}

\author{Xingbo~Zhao}
\affiliation{Institute of Modern Physics, Chinese Academy of Sciences, Lanzhou 730000, China}
\affiliation{School of Nuclear Science and Technology, University of Chinese Academy of Sciences, Beijing 100049, China}
\affiliation{CAS Key Laboratory of High Precision Nuclear Spectroscopy, Institute of Modern Physics, Chinese Academy of Sciences, Lanzhou 730000, China}

\author{Vladimir~A.~Karmanov}
\affiliation{Lebedev Physical Institute, Leninsky Prospekt 53, 119991 Moscow, Russia}

\author{James~P. Vary}
\affiliation{Department of Physics and Astronomy, Iowa State University, Ames, Iowa 50011}

\date{\today}

\begin{abstract}
 
 We compute the gravitational form factors $A_i$, $D_i$ and $\bar c_i$ of the scalar Yukawa theory using both the light-cone and covariant perturbation theory at the one-loop level. 
 The light-cone formalism provides a potential approach to access these form factors beyond the perturbative regime. However, 
unlike the covariant formulation, the Poincaré symmetry on the light cone is not manifest. In this work, we use perturbation theory as a benchmark to extract the gravitational form factors from the light-front energy-momentum tensor. 
By comparing results on and off the light cone, we identify $T^{++}, T^{+a}, T^{+-}, T^{12}$ as the ``good currents" that are properly renormalized and can be used to extract the gravitational form factors. 
\end{abstract}

 \maketitle 
 
 \section{Introduction}\label{sect:introduction}
 
 The energy-momentum tensor (EMT) $T^{\mu\nu}$ is an observable of fundamental significance. It encodes the internal energy, momentum, angular momentum and stress distributions of a system \cite{Teryaev:2016edw}. Therefore, this quantity is particularly interesting for unraveling the structures of hadrons, quarks and gluons bound together by the strong force \cite{Polyakov:2018zvc,Burkert:2023wzr}. 
 The hadronic matrix element (HME) of the EMT can be parametrized in terms of Lorentz tensors \cite{Kobzarev:1962wt, Pagels:1966zza}. The scalar functions that are specific to each hadron are known as the gravitational form factors (GFFs). These quantities then probe the internal dynamics of the hadron \cite{Ji:1995cu, Ji:1996ek}. 
 
 Since QCD is strongly-coupled at the hadron scale, non-perturbative methods are needed to access the GFFs. Among various approaches, the light-front Hamiltonian formulation is poised as a natural framework to describe hadrons as  relativistic self-bound states \cite{Namyslowski:1985zq, Zhang:1994ti, Burkardt:1995ct, Brodsky:1997de, Carbonell:1998rj, Heinzl:2000ht, Miller:2000kv, Bakker:2013cea, Hiller:2016itl, Brodsky:2022fqy, Gross:2022hyw}. 
 This approach is demonstrated in one of our recent works for a strongly-coupled scalar Yukawa theory \cite{Cao:2023ohj}. In that work, we extracted GFFs $A(Q^2)$ and $D(Q^2)$ from $T^{++}$ and $T^{+-}$, two tensor components of the EMT. We showed that these two components can be renormalized using the same counterterms obtained from solving the light-front Schrödinger equation, i.e. the renormalization of the light-front Hamiltonian \cite{Li:2014kfa, Li:2015iaw, Li:2015zah, Karmanov:2016yzu}. 
 
Further quantities of interest are the GFFs associated with each constituent, which allow us to decompose a hadronic density into contributions from each constituent, e.g. quarks and gluons \cite{Ji:1994av, Ji:1995sv, Lorce:2018egm, Lorce:2021xku}. For the scalar theory, apart from GFFs $A_i(Q^2)$ and $D_i(Q^2)$, where $i$ enumerates over the constituents, there is an additional GFF, $\bar c_i(Q^2)$ \cite{Polyakov:2018zvc}.  It represents the force between constituent $i$ and the rest of the system. Current conservation $\partial_\mu T^{\mu\nu} = 0$ implies that the total $\bar c$ vanishes, 
\begin{equation}
\bar c(Q^2) \equiv \sum_i \bar c_i (Q^2)= 0\,.
\end{equation}
Namely, all inter-particle forces balance out for an isolated system -- the quantum version of Newton's third law. 
 To extract $\bar c_i(Q^2)$, using the above two currents $T_i^{++}$ and $T^{+-}_i$ are not sufficient, and we need additional currents such as $T^{12}_i, T^{11}_i, T^{22}_i$. However, these operators contain uncanceled divergences and lead to a violation of the current conservation. These difficulties are caused by the breaking of the Poincaré symmetry due to truncation within the light-front formalism. This effect is better encapsulated in the explicitly covariant formulation of light-front dynamics (CLFD) (see Ref.~\cite{Carbonell:1998rj} for a review). In CLFD, $T^{12}, T^{11}, T^{22}$ are contaminated by the spurious contributions that depend on the orientation of the light front \cite{Cao:2023ohj}. The most general Lorentz structures of the EMT taking into account CLFD are given by Lorcé \cite{Lorce:2015lna}.
 
 Facing these difficulties in the non-perturbative calculation, we turn to perturbation theory for guidance. 
Recently, More~et.~al. evaluated the EMT of a quark dressed by gluons using leading order (LO) light-cone perturbation theory (LCPT) \cite{More:2021stk,More:2023pcy}. Their results can be compared with the recent LO QED calculations in covariant perturbation theory (CPT) \cite{Metz:2021lqv, Freese:2022ibw, Freese:2022jlu, Eides:2023uox}. More~et.~al. suggested to use spin-flip HMEs to extract GFFs $D_i(Q^2)$ and $\bar c_i(Q^2)$, which are unfortunately not available in the scalar case. Furthermore, they obtained a non-vanishing $\bar c (Q^2)$ except at $Q = 0$. They attributed the violation of current conservation to zero modes \cite{More:2023pcy}. 
In this work, we compute the EMT of the scalar Yukawa theory using both covariant and LCPT at the one-loop level. 
Scalar theory is among the simplest yet non-trivial quantum field theories in 3+1 dimensions \cite{Baym:1960zz, Nieuwenhuis:1996mc, Gross:2001ha, Hwang:2004mr, Emami-Razavi:2010pmi, Emami-Razavi:2011bxe, Ji:2012qd, Carbonell:2012bv, Rochev:2013gfp, Gigante:2016dtz, Chabysheva:2021yse, Chabysheva:2022duu, Cao:2023ohj, Bray:2023eug}. The zero modes within the scalar theory are only constrained degrees of freedom \cite{Harindranath:1999vf, Hiller:2016itl}. Therefore, it is an ideal testbed for developing computational techniques in light-front field theory. 
By comparing the covariant and light-cone perturbation theory for the scalar Yukawa model, we identify the sources of the divergences and explore how to properly extract GFFs $A_i(Q^2)$, $D_i(Q^2)$ and $\bar c_i(Q^2)$ from the light-front EMT.  

The remaining part of the article is organized as follows. We introduce the scalar Yukawa model including the one-loop renormalization of this theory in Sect.~\ref{sect:formalism}. We then compute the energy-momentum tensor up to one-loop in both the covariant perturbation theory and in light-cone perturbation theory in Sect.~\ref{sect:emt}. We analyze these results using covariant light-front dynamics in Sect.~\ref{sect:clfd}. Finally, we summarize in Sect.~\ref{sect:summary}. 


\section{Scalar Yukawa theory}\label{sect:formalism}
 
 In this work, we consider the scalar Yukawa theory. The classical Lagrangian density of this model reads \cite{Li:2014kfa, Li:2015iaw, Li:2015zah, Karmanov:2016yzu}, 
 \begin{equation}
 \mathscr L = \partial_\mu N^\dagger \partial^\mu N - m^2 N^\dagger N \\
 + \frac{1}{2}\partial_\mu \pi \partial^\mu \pi - \frac{1}{2}\mu^2 \pi^2 + g N^\dagger N \pi\,.
 \end{equation}
 This Lagrangian describes a complex scalar field $N$ interacting with a real scalar field $\pi$ via a Yukawa type coupling. It can be used as a rudimentary model for the nucleon-pion interaction. Therefore, we will refer to the complex scalar field $N$ as the (mock) nucleon field and tentatively assign the mass of the nucleon $m = 0.94\,\mathrm{GeV}$ to its physical mass. Similarly, the real scalar field $\pi$ will be referred to as the (mock) pion field, and tentatively assigned the pion mass $\mu = 0.14\,\mathrm{GeV}$.  The coupling $g$ is dimensionful. Hence, this model is super-renormalizable. However, there are still divergences within the one-loop radiative correction. It is useful to introduce a dimensionless coupling $\alpha = g^2/(16\pi m^2)$. This coupling is the Yukawa coupling appearing in the tree-level Yukawa potential, 
 \begin{equation}
 V(r) = - \frac{\alpha}{r}e^{-\mu r}\,.
 \end{equation}
 
Specifically, we consider the nucleon dressed by the pion up to order $O(\alpha)$. 
We adopt the bare renormalization scheme \cite{Peskin:1995ev}, i.e. the field operator is not normalized, 
$\langle 0 | N^\dagger(x) | p\rangle = \sqrt{Z} e^{ip\cdot x}$. 
Instead, the state vector will be normalized to unity, $\langle p' | p \rangle = 2p^+(2\pi)^3\delta^3(p-p')$. 
The normalization of the state vector is easier to be generalized to the non-perturbative regime in the light-front Hamiltonian formulation \cite{Karmanov:2016yzu}.  
Within this scheme, the Lagrangian can be written as, 
 \begin{equation}
 \mathscr L = \partial_\mu N^\dagger \partial^\mu N - m^2 N^\dagger N \\
 + \frac{1}{2}\partial_\mu \pi \partial^\mu \pi - \frac{1}{2}\mu^2 \pi^2 + g_0 N^\dagger N \pi + \delta m^2 N^\dagger N
 + \frac{1}{2} \delta \mu^2 \pi^2 \,,
 \end{equation}
 where, $\delta m^2$ and $\delta \mu^2$ are mass counterterms and $g_0$ is the bare coupling. These parameters can be obtained from the mass and coupling constant renormalization. 
 
\begin{figure}
\centering
\includegraphics[width=0.2\textwidth]{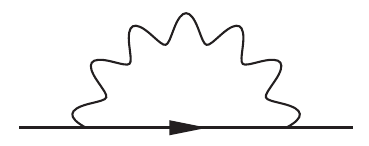}
\caption{Leading order diagram for the self-energy of the scalar Yukawa model}
\label{fig:self-energy}
\end{figure}
 
 We adopt two types of regularization schemes, the dimensional regularization (DR) and Pauli-Villars regularization (PV). The former is the standard choice in covariant perturbation theory and the latter is useful for both the covariant and light-cone perturbation theory \cite{Brodsky:2001ja, Brodsky:2002tp, Brodsky:2005yu, Chabysheva:2009aw, Chabysheva:2009vm, Chabysheva:2010vk}.  Then the nucleon self-energy up to the LO order (see Fig.~\ref{fig:self-energy}) is, 
 \begin{equation}\label{eqn:self-energy}
 \Sigma(p^2) = -\frac{g_0^2}{16\pi^2} 
 \begin{dcases*}
 \frac{1}{\epsilon} + \ln \frac{4\pi}{e^{\gamma_E}} - \int_0^1 \dd x \ln \big[(1-x)\mu^2 + x m^2 - x(1-x)p^2 \big] , & \text{DR} \\
 \int_0^1 \dd x \ln \frac{(1-x)\mu^2_\textsc{pv} + x m^2 - x(1-x)p^2}{(1-x)\mu^2 + x m^2 - x(1-x)p^2} , & \text{PV}
 \end{dcases*}
 \end{equation}
 where, $\epsilon = 2 - d/2$, and $d$ is the spacetime dimension. $\gamma_E \approx 0.577216$ is the Euler's constant. 
With the on-shell scheme, the mass counterterm is then $\delta m^2 = \Sigma(m^2)$. The $Z$-factor is, 
\begin{equation}\label{eqn:Z}
Z = \Big[ 1 - \frac{\dd \Sigma(p^2)}{\dd p^2} \Big]^{-1}\Big|_{p^2 \to m^2}.
\end{equation}
Note that $Z$ is not divergent in the scalar Yukawa model. 
The bare coupling is obtained from imposing the physical coupling $g$ to the nucleon-pion scattering amplitude at the on-shell point. And the result, according to the Lehmann-Symanzik-Zimmermann reduction formula \cite{Peskin:1995ev}, is, 
\begin{equation}\label{eqn:bare_coupling}
g_0 = \frac{g}{Z\sqrt{Z'}}\,,
\end{equation}
where $Z'$ is the field renormalization constant of the pion, 
\begin{equation}\label{eqn:Zprime}
Z' = \Big[1 + \frac{g_0^2}{(4\pi)^2}\int_0^1 dx \frac{x(1-x)}{m^2 - x(1-x)\mu^2}\Big]^{-1}\,.
\end{equation}

\section{Energy-momentum tensor}\label{sect:emt}

 The EMT of the scalar Yukawa theory is, 
 \begin{multline}
 T^{\mu\nu} = \partial^{\{\mu}N^\dagger \partial^{\nu\}}N - g^{\mu\nu} (\partial_\rho N^\dagger \partial^\rho N - m^2 N^\dagger N) 
 - g^{\mu\nu} g_0 N^\dagger N \pi - g^{\mu\nu} \delta m^2 N^\dagger N \\
 + \partial^\mu \pi \partial^\nu \pi - \frac{1}{2} g^{\mu\nu} (\partial_\sigma \pi \partial^\sigma \pi - \mu^2 \pi^2) 
 - \frac{1}{2}g^{\mu\nu} \delta \mu^2 \pi^2\,.
 \end{multline}
 Here, $a^{\{\mu}b^{\nu\}} \equiv a^\mu b^\nu + a^\nu b^\mu$. This EMT is symmetric, and there is no distinction between the canonical EMT and the Belinfante–Rosenfeld-Hilbert-Einstein EMT. 
 In analogy to mass decomposition in QCD \cite{Ji:1994av, Ji:1995sv}, we split the EMT
into two parts, 
\begin{align}
T^{\mu\nu}_N =\,& \partial^{\{\mu}N^\dagger \partial^{\nu\}}N - g^{\mu\nu} (\partial_\rho N^\dagger \partial^\rho N - m^2 N^\dagger N) \nonumber  - g^{\mu\nu} g N^\dagger N \pi - g^{\mu\nu}\delta m^2 N^\dagger N\,, \\
T^{\mu\nu}_\pi =\,& \partial^\mu \pi \partial^\nu \pi - \frac{1}{2} g^{\mu\nu} (\partial_\sigma \pi \partial^\sigma \pi - \mu^2 \pi^2) - \frac{1}{2}g^{\mu\nu}\delta \mu^2 \pi^2\,,
\end{align}
i.e. the nucleon EMT contains the interaction. 
The HMEs of the EMT for a spin-0 particle can be parametrized by the following Lorentz tensors \cite{Polyakov:2018zvc},
\begin{equation}\label{eqn:HME_EMT}
    t^{\alpha\beta}_i = \langle p' | T^{\alpha\beta}_i(0)|p\rangle = 2P^\alpha P^\beta A_i(-q^2) 
+ \frac{1}{2}\big(q^\alpha q^\beta - g^{\alpha\beta} q^2\big) D_i(-q^2) + 2M^2 g^{\alpha\beta} \bar c_i (-q^2)\,.
\end{equation} 
Here, $P = (p'+p)/2$, $q = p' - p$. $M$ is the mass of the scalar, i.e. $p^2 = p'^2 = M^2$. The Lorentz scalar functions $A_i(-q^2)$, $D_i(-q^2)$ and $\bar c_i(-q^2)$ are the GFFs that encode the internal dynamics of the physical particle. The subscript $i = N, \pi$ enumerates over the constituents. Summing over all constituents gives the total EMT. Accordingly, 
\begin{align}
\sum_i A_i(-q^2) =\,& A(-q^2), \\
\sum_i D_i(-q^2) =\,& D(-q^2), \\
\sum_i \bar c_i(-q^2) =\,& \bar c(-q^2)\,.
\end{align} 
Current conservation 
\begin{equation}
\partial_\nu T^{\mu\nu} = 0\,,
\end{equation}
 implies 
 \begin{equation}\label{eqn:c-bar_sumrule}
 \bar c(-q^2) = 0\,.
 \end{equation}
 Physically, $\bar c_i$ represents the force between the $i$-th constituent and the rest of the system. Eq.~(\ref{eqn:c-bar_sumrule}) is nothing but Newton's third law applied to hadrons: the total force vanishes for a self-bound system. Current conservation also protects the EMT $T^{\alpha\beta}$ against the quantum fluctuation. Consequently, the GFFs $A(-q^2)$ and $D(-q^2)$ do not depend on the renormalization scheme or scale. By contrast, the GFFs $A_i(-q^2)$, $D_i(-q^2)$ and $\bar c_i(-q^2)$ for each constituent are renormalization scheme and scale dependent. Unless otherwise stated, these dependences are suppressed for simplicity. Despite this arbitrariness, the decomposition of the EMT provides insights into the system. For example, in QCD, mass decomposition based on the EMT provides answers to the origin of the proton mass in terms of quark mass, quark kinetic energy, gluon kinetic energy and the anomalous mass \cite{Ji:1994av, Ji:1995sv, Lorce:2018egm, Lorce:2021xku}.

\begin{figure}
\centering
\includegraphics[width=0.45\textwidth]{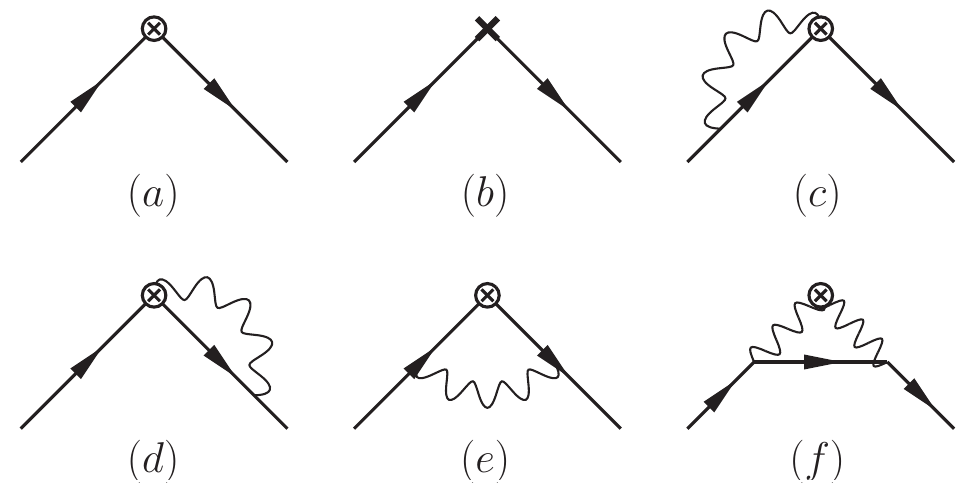}
\caption{Leading-order diagrams for the energy-momentum tensor. The solid lines with an arrow represent the nucleon. The wavy lines represent the pion. The cross represents the mass counterterm. The circled crosses represent the operator insertion. }
\label{fig:1-loop_diagrams}
\end{figure}

Figure \ref{fig:1-loop_diagrams} shows the LO diagrams for the EMT. Diagrams ($a$)--($e$) contribute to $T^{\mu\nu}_N$ while diagram ($f$) is the contribution to $T^{\mu\nu}_\pi$. We first evaluate these diagrams in covariant perturbation theory and extract the relevant GFFs. These results will serve as a benchmark for the LCPT. 
 
 \subsection{Covariant perturbation theory}\label{sect:CPT}
 
The covariant calculation is straightforward. The results are, 
 \begin{align}
 t^{\mu\nu}_a =\,& Z\Big[2P^\mu P^\nu - \frac{1}{2}(q^\mu q^\nu - q^2 g^{\mu\nu}) \Big], \\
 t^{\mu\nu}_b =\,& - t^{\mu\nu}_c = - t^{\mu\nu}_d = - Zg^{\mu\nu} \Sigma(m^2), \\
 t^{\mu\nu}_e =\,& Z g_0^2 \int_0^1 \dd x \int_0^{1-x} \dd y \int \frac{\dd^4 \ell_E}{(2\pi)^4} \frac{\mathcal O^{\mu\nu}_e}{(\ell_E^2 + \Delta_e)^3}, \\
 t^{\mu\nu}_f =\,& Z g_0^2 \int_0^1 \dd x \int_0^{1-x} \dd y \int \frac{\dd^4 \ell_E}{(2\pi)^4} \frac{\mathcal O^{\mu\nu}_f}{(\ell_E^2 + \Delta_f)^3}\,,
\end{align}
where, $\Delta_e = (x+y)^2 m^2+(1-x-y)\mu^2 - xy q^2$, $\Delta_f = (1-x-y)^2 m^2 + (x+y)\mu^2 - xy q^2$ and 
$\mathcal O^{\mu\nu}_e = g^{\mu\nu} \ell^2_E + 2(1-x-y)^2 2P^\mu P^\nu 
+ ((x-y)^2-1) (q^\mu q^\nu - q^2 g^{\mu\nu}) + 2 g^{\mu\nu} \Big\{ \big[1 - (1-x-y)^2\big] m^2 + \frac{1}{2}(x^2-x+y^2-y) q^2\Big\}$, 
$\mathcal O^{\mu\nu}_f = g^{\mu\nu} \ell^2_E + 2(1-x-y)^2 2P^\mu P^\nu + \big[(x-y)^2 -1 \big] (q^\mu q^\nu - g^{\mu\nu}q^2) + 2g^{\mu\nu} \big\{\mu^2 - (1-x-y)^2 m^2 + \frac{1}{2}(x^2-x+y^2-y) q^2\big\}$.

From these HMEs, we can extract the GFFs. It turned out, GFFs $A_i$ and $D_i$ are divergence free, 
\begin{align}
A_N(Q^2) =\,& Z+ Z\frac{g_0^2}{(4\pi)^2} \int_0^1 \dd x \int_0^{1-x}\dd y \frac{(1-x-y)^2}{\Delta_e}, \\
A_\pi(Q^2) =\,& Z\frac{g_0^2}{(4\pi)^2} \int_0^1 \dd x \int_0^{1-x}\dd y \frac{(1-x-y)^2}{\Delta_f}, \\
D_N(Q^2) =\,& - Z - Z\frac{g_0^2}{(4\pi)^2} \int_0^1 \dd x \int_0^{1-x}\dd y \frac{1-(x-y)^2}{\Delta_e}, \\
D_\pi(Q^2) =\,& - Z\frac{g_0^2}{(4\pi)^2} \int_0^1 \dd x \int_0^{1-x}\dd y \frac{1-(x-y)^2}{\Delta_f}\,.
\end{align}
On the other hand, GFFs $\bar c_i$ are UV divergent. The results with dimensional regularization are, 
\begin{multline}
    \bar{c}_{N}^{(\overline{\textsc{ms}})}(Q^2, \mu^2_R) = - \frac{Z g_0^2}{64\pi^2m^2} \Big(\frac{1}{\epsilon}+1\Big)  + \frac{Z g_0^2}{32\pi^2 m^2} \int_0^1 \dd x \ln \frac{(1-x)\mu^2 + x^2m^2}{ \mu^2_R}  - \frac{Z g_0^2}{32\pi^2m^2} \int_0^1 \dd x \int_0^{1-x} \dd y \ln \frac{\Delta_e}{ \mu^2_R} \\ 
    + \frac{Z g_0^2}{64\pi^2m^2} \int_0^1 \dd x \int_0^{1-x} \dd y \frac{2(x+y)(2-x-y)m^2 + (x^2-x+y^2-y) q^2}{\Delta_e},
    \label{eq:cBarCovDimNuc} 
\end{multline}
and,
\begin{multline}
    \bar{c}_{\pi}^{(\overline{\textsc{ms}})}(Q^2, \mu_R^2) =  \frac{Z g_0^2}{64\pi^2m^2} \Big(\frac{1}{\epsilon}-1\Big) - \frac{Z g_0^2}{32\pi^2m^2} \int_0^1 \dd x \int_0^{1-x}\dd y \ln \frac{\Delta_f}{{\mu}^2_R}\\
    + \frac{Z g_0^2}{64\pi^2m^2} \int_0^1 \dd x \int\limits_0^{1-x} \dd y \frac{2\mu^2 - 2(1-x-y)^2m^2 + (x^2-x+y^2-y) q^2 }{\Delta_f}.
    \label{eq:cBarCovDimPion}
\end{multline}
Here, $ \mu_R$ is the renormalization scale in the $\overline{\mathrm{MS}}$ scheme. 
The GFFs $\bar c_i$ can also be obtained with the Pauli-Villars regularization. The results are,
\begin{multline}
    \bar{c}_{N}^{(\textsc{pv})}(Q^2, \mu_{\textsc{pv}}^2) = - \frac{Z g_0^2}{32\pi^2m^2}\int_0^1 \dd x \ln \frac{(1-x)\mu^2_\textsc{pv} + x^2m^2}{(1-x)\mu^2 + x^2m^2} + \frac{Z g_0^2}{32\pi^2m^2} \int_0^1 \dd x \int_0^{1-x} \dd y  \ln \frac{\Delta_{e}^{(1)}}{\Delta_e}\\
    + \frac{Z g_0^2}{64\pi^2m^2} \int_0^1 \dd x \int_0^{1-x} \dd y \sum_j(-1)^j \frac{2(x+y)(2-x-y)m^2 + (x^2-x+y^2-y)q^2}{\Delta_{e}^{(j)}},
    \label{eq:cBarCovPVNuc}
\end{multline}
Here, the index ``$j$'' runs over 0 and 1 for the original field and the Pauli-Villars field respectively.
And,
\begin{multline}
    \bar{c}_{\pi}^{(\textsc{pv})}(Q^2, \mu_{\textsc{pv}}^2) = \frac{Z g_0^2}{32\pi^2m^2} \int_0^1 \dd x \int_0^{1-x} \dd y  \ln \frac{\Delta_{f}^{(1)}}{\Delta_f}\\
    + \frac{Z g_0^2}{64\pi^2m^2} \int_0^1 \dd x \int_0^{1-x} \dd y \sum_j(-1)^j \frac{2\mu_j^2-2(1-x-y)^2m^2+(x^2-x+y^2-y)q^2}{\Delta_{f}^{(j)}}.
    \label{eq:cBarCovPVPion}
\end{multline}
In these expressions, $\Delta_e^{(0)} = \Delta _e$, and $\Delta_e^{(1)}$ is obtained from $\Delta _e$ by replacing $\mu$ with $\mu_1 = \mu_\textsc{pv}$. Similarly,  $\Delta_f^{(0)} = \Delta _f$, and $\Delta_f^{(1)}$ is obtained from $\Delta _f$ by replacing $\mu$ with $\mu_1 = \mu_\textsc{pv}$. 

The obtained $\bar c_i(Q^2)$ for both the nucleon and the pion is constant, i.e. independent of $Q^2$, consistent with the ultra-local character of the scalar Yukawa theory.  
The total $\bar c(Q^2) = \bar c_N(Q^2) + \bar c_\pi(Q^2)$, including the divergences, vanishes with both the dimensional (DR) and Pauli-Villars (PV) regularizations,  as required by current conservation. 

Comparing the results from the dimensional regularization and the Pauli-Villars regularization, it is readily established that the Pauli-Villars results can be obtained from the dimensional regularization with the following replacement, 
\begin{equation}
\frac{1}{\epsilon} \; \leftrightarrow \; \ln \frac{\mu^2_{\textsc{pv}}}{\mu_R^2} - \frac{3}{2},
\end{equation}
in the physical limit ($\mu_{\textsc{pv}} \to \infty$ and $\epsilon \to 0$).

\subsection{Light-cone perturbation theory}\label{sect:LCPT}

We adopt the Pauli-Villars regularization for calculations in LCPT \cite{Brodsky:2001ja, Brodsky:2002tp, Brodsky:2005yu}. This regularization scheme can be extended to the non-perturbative light-front Hamiltonian formalism including gauge theories \cite{Chabysheva:2009aw, Chabysheva:2009vm, Chabysheva:2010vk, Chabysheva:2015vga}. In LO LCPT, the obtained counterterms are the same as the counterterms in covariant perturbation theory with Pauli-Villars regularization, i.e. Eqs.~(\ref{eqn:self-energy}--\ref{eqn:Zprime}). 

For evaluating the HMEs of the EMT, we adopt the Drell-Yan frame, in which the longitudinal momentum of the graviton vanishes $q^+ = 0$ \cite{Drell:1969km}. The vacuum creation and annihilation are suppressed in this frame. Mathematically, $q^+=0$ is equivalent to integrating out the longitudinal $x^-$ direction on the light cone.  In the Drell-Yan frame, the diagrams needed for the EMT are the same as Fig.~\ref{fig:1-loop_diagrams}, except all diagrams are now ordered in light-front time $x^+$.  Applying the diagrammatic rules \cite{Brodsky:1997de}, we obtain the HMEs as, 
\begin{align}
    t_{a}^{\mu\nu} =\,& Z \left[ 2P^\mu P^\nu - \frac{1}{2}(q^\mu q^\nu-q^2 g^{\mu\nu}) \right], \\
    t_{b }^{\mu\nu} =\,&  -t_c^{\mu\nu} = -t_d^{\mu\nu} = -Z g^{\mu\nu} \Sigma(m^2), \\
    t_{e}^{\mu\nu} =\,& Z g_0^2 \int_0^1 \frac{\dd x}{2x(1-x)}\int \frac{\dd^2k_\perp}{(2\pi)^3} \sum_j(-1)^j \frac{\frac{{1}}{2}q^2 g^{\mu\nu}+p_2^{\{\mu}p_2^{\prime\nu\}}}{(1-x)(s_2^j-m^2)(s_{2a}^j-m^2)}, \\
    t_f^{\mu\nu} =\,& Z g_0^2 \int_0^1 \frac{\dd x}{2x(1-x)} \int \frac{\dd^2k_\perp}{(2\pi)^3} \sum_j(-1)^j \frac{\frac{1}{2}q^2 g^{\mu\nu} + p_{1j}^{\{\mu}p_{1j}^{\prime\nu\}}}{x(s^j_2-m^2)(s^j_{2b}-m^2)},
\end{align}
where, %
\begin{align}
p^\mu_{1j} =\,& \big(xP^+, \frac{\vec{p}_{1j\perp}^2+\mu^2_j}{p_{1j}^+}, \vec{k}_\perp + x\vec{P}_\perp - \frac{1}{2}x\vec{q}_\perp\big), \\
p'^\mu_{1j} =\,& \big(xP^+,\frac{\vec{p}_{1j\perp}^{\prime2}+\mu^2_j}{p_{1j}^{\prime+}}, \vec{k}_\perp + x\vec{P}_\perp + (1 - \frac{1}{2}x)\vec{q}_\perp \big), \\
p^\mu_2 = \,& \big((1-x)P^+, \frac{\vec{p}_{2\perp}^2+m^2}{p_2^+},  -\vec{k}_\perp + (1-x)\vec{P}_\perp - \frac{1}{2}(1-x)\vec{q}_\perp \big), \\
p'^\mu_2 =\,& \big((1-x)P^+, \frac{\vec{p}_{2\perp}^{\prime2}+m^2}{p_2^{\prime+}}, -\vec{k}_\perp + (1-x)\vec{P}_\perp + \frac{1}{2}(1+x)\vec{q}_\perp\big).
\end{align}
 The $s$ factors in the energy denominators are, 
\begin{align}
    s_2^j & =  \frac{\vec{k}_\perp^2+\mu_j^2}{x} + \frac{\vec{k}_\perp^2+m^2}{1-x}, \\
    s_{2a}^j &= \frac{(\vec{k}_\perp-x\vec{q}_\perp)^2+\mu_j^2}{x} + \frac{(\vec{k}_\perp-x\vec{q}_\perp)^2+m^2}{1-x}, \\
    s^j_{2b} &= \frac{[\vec{k}_\perp+(1-x)\vec{q}_\perp]^2+\mu_j^2}{x} + \frac{[\vec{k}_\perp+(1-x)\vec{q}_\perp]^2+m^2}{1-x}.
\end{align}
As mentioned, in Pauli-Villars regularization, we introduce $\mu_0 = \mu$ and $\mu_1 = \mu_\textsc{pv}$. 

From these expressions, it is clear that the light-front results are not manifestly covariant and different components have to be evaluated separately. In our previous work, we proposed to use $t^{++}_i$ to extract GFF $A_i$ and to use $t^{++}_i, t^{+-}_i$ to extract GFF $D_i$ \cite{Cao:2023ohj, Xu:2024cfa}. Using the perturbative results obtained above, one readily shows that $t^{++}_i$ and $t^{+-}_i$ from LCPT and from the covariant perturbation theory are identical diagram by diagram, which validates our previous proposal in the context of  perturbation theory. 
Unfortunately, this equivalence could not be extended to all components of the EMT. Indeed, one can check that current is not conserved, viz. 
\begin{equation}
q_\nu t^{1\nu} \ne 0\,.
\end{equation}
From the standard Lorentz decomposition Eq.~(\ref{eqn:HME_EMT}), since individually, 
\begin{equation}
q_\nu t^{1\nu}_i = 2m^2 q_\perp^1 \bar c_i(-q^2)\,,
\end{equation}
the violation of the current conservation leads to a divergent total $\bar c$:
\begin{equation}
\bar c(-q^2) \equiv \sum_i \bar c_i(-q^2) = \frac{Z}{m^2} \Sigma(m^2),
\end{equation}
 in the standard formulation (\ref{eqn:HME_EMT}). 
Fortunately, we can compare with the covariant results diagram by diagram to pinpoint the sources of the violation. 
We find that the violation of the current conservation originates from $t^{11}_i, t^{22}_i$ with contributions from diagrams ($e$) and ($f$). The origin of the discrepancy is that the counterterms derived from renormalizing the Hamiltonian operator are not sufficient to remove the divergences in $t^{11}, t^{22}$. This is in contrast to $t^{+-}$ which in fact involves the same operator as the Hamiltonian. 
On the other hand, $t^{+1}_i, t^{+2}_i, t^{12}_i$ are identical to the respective covariant results.  %
The minus HMEs, $t^{-i}_i, t^{--}_i$ are very singular and cannot be regularized with a single Pauli-Villars pion. More copies of the Pauli-Villars ghosts are needed to tame these components \cite{Bray:2023eug}. 

In summary, $t^{++}_i, t^{+-}_i, t^{+1}_i, t^{+2}_i, t^{12}_i$ on the light front are identical to their counterparts in covariant perturbation theory, while $t^{11}_i, t^{22}_i$ are not. $t^{11}_i, t^{22}_i$ also contain uncanceled divergences. We can use these ``good currents" $t^{++}_i, t^{+-}_i, t^{+1}_i, t^{+2}_i, t^{12}_i$ to extract the GFFs, 
\begin{align}  
    t_i^{++} &= 2(P^+)^2 A_i(Q^2), \label{eqn:t++}\\
    t_i^{+a} &= 2P^+P^a_\perp A_i(Q^2) + \frac{1}{2}q^+ q_\perp^a D_i(Q^2),  \label{eqn:t+a}\\
    t_i^{12} &= 2P^1_\perp P^2_\perp A_i(Q^2) + \frac{1}{2}q^1_\perp q_\perp^2 D_i(Q^2), \label{eqn:t12}\\
    t_i^{+-} &= 2\big(P^2_\perp + M^2 + \frac{1}{4}q^2_\perp\big) A_i(Q^2) + {q}_\perp^2 D_i(Q^2) + 4M^2 \bar{c}_i(Q^2), \label{eqn:t+-}
\end{align}
where, $a=1,2$, $i = N, \pi$ and $Q^2 = -q^2$. We can use the combinations of these HMEs to extract the GFFs $A_i, D_i, \bar c_i$. The above expressions are further simplified in the transverse Breit frame with $\vec P_\perp = 0$. 
Note that to use $t^{+a}$ to extract $D_i$, we need to compute the EMT in a general frame $q^+\ne 0$ and take the Drell-Yan limit $q^+ \to 0$ in the end. Since $\bar c_i$ is only contained in $t^{+-}_i$ for the above set of currents, this HME is indispensable to obtain the forces between the nucleon and the pion. 

\section{Covariant light-front dynamics}\label{sect:clfd}

The conventional wisdom regarding currents in light-front physics is that one should use the ``good currents" while avoiding the ``bad" and ``terrible currents" if possible \cite{Burkardt:1995ct, Bakker:2002aw, Brodsky:1998hn, Choi:2004ww, Li:2017uug, Jaus:1989au, Jaus:1999zv, Arifi:2023uqc,Tang:2020org, Li:2018uif, Krutov:2018mbu, Strikman:2017koc, Choi:2005fj, Cooke:2001kz, Krutov:2001gu, Choi:1998jd, Melnitchouk:1996vp, Bolz:1996sw, Burkardt:1996gp, Burkardt:2000za, Vanderhaeghen:1999xj}. The good currents are those associated with the plus component of the current, e.g. $J^+$. The bad currents are currents with transverse components, i.e. $\vec J_\perp$. And the terrible (or worst) currents are those associated with the minus component of the current, e.g. $J^-$ \cite{Strikman:2017koc}. Quantitatively, one can define $\gamma$ the ``goodness" of an operator $O$ as its scaling dimension under the longitudinal boost, viz \cite{Leutwyler:1977vy}.
\begin{equation}
\big[K_3, O\big] = -i \gamma O,
\end{equation}
where, $K_3$ is the  generator of the longitudinal boost. From the infinite-momentum perspective, good currents are those that increase as $P_z$ increases, i.e. relevant. Bad and terrible currents are those that decrease as $P_z$ increases \cite{Frankfurt:1988nt, Strikman:2017koc}. Our LCPT calculation certainly follows the conventional wisdom. On the other hand, the traditional good currents $t^{++}$ (goodness 2) and $t^{+a}$ (goodness 1) are not sufficient to extract the full set of GFFs. And the goodness of the currents does not distinguish the UV divergent $t^{11}, t^{22}$ from the UV finite $t^{+-}$, $t^{12}$, all of which have goodness 0. 

A systematic way to analyze the current components based on symmetry is the covariant light-front dynamics proposed by Karmanov et. al \cite{Close:1971bp, Osborn:1972dy, Berestetsky:1977zk, Karmanov:1991fv, Karmanov:1996un, Carbonell:1998rj, Carbonell:1999pt, Carbonell:2000rr, Karmanov:2002qu, Brodsky:2003pw}. The rationale of this method is as follows. The Lorentz decomposition of the HME is based on the Poincaré symmetry.  In light-front dynamics, some of the Poincaré symmetries are not manifest \cite{Dirac:1949cp}. In practical calculations, these symmetries are likely broken. As a result, the HMEs of the EMT have to be re-parametrized with the reduced symmetries to account for the possible breaking of the Poincaré symmetries. And the full Poincaré symmetry is restored in the continuum limit. An effective way to parametrize the HME with the reduced symmetries is to include a null vector $\omega^\mu$ ($\omega_\mu \omega^\mu = 0$), which is perpendicular to the light front, i.e. $\omega\cdot x = 0$ \cite{Karmanov:1991fv, Karmanov:1996un, Carbonell:1998rj, Carbonell:1999pt, Carbonell:2000rr, Karmanov:2002qu, Brodsky:2003pw}. Standard light-front coordinates correspond to the choice $\omega^\mu = (\omega^+, \omega^-, \vec\omega_\perp) = (0, 2, 0)$. There is a conformal symmetry related to the scaling of $\omega$. Namely, physics does not change if we multiply $\omega$ with an arbitrary non-zero real number. This symmetry requires that the $\omega^\mu$ dependence always appears in the form $\omega^\mu/(\omega\cdot P)$ \cite{Carbonell:1998rj}. 

The presence of $\omega$ also adds one more scalar $\zeta = (\omega\cdot q)/(\omega\cdot P)$ into the calculation. Now, the form factors depend on two Lorentz scalars, $q^2$ and $\zeta$. It can be shown that Hermiticity of the EMT operator leads to $F_i(\zeta, q^2) = F^*_i(-\zeta, q^2)$, where $F_i$ is a form factor \cite{Lorce:2015lna}. In the Drell-Yan frame $\zeta = 0$ and the form factors become real functions that depend on $q^2$ only. This choice dramatically simplifies the Lorentz decomposition in covariant light-front dynamics \cite{Carbonell:1999pt}. We will adopt the Drell-Yan frame unless otherwise stated. 
Then, one of the most general Lorentz decompositions of the HMEs of the EMT operator for a spin-0 hadron on the light front in the Drell-Yan frame $q^+ = 0$ is\footnote{This decomposition differs from our previous work \cite{Cao:2023ohj} by replacing tensor $V^\alpha V^\beta$ with $V^\alpha V^\beta + q^\alpha q^\beta$. The latter violates current conservation. Another minor difference is to replace ${q^4 \omega^\alpha \omega^\beta}/{(\omega\cdot P)^2}$ with ${M^4 \omega^\alpha \omega^\beta}/{(\omega\cdot P)^2}$. The physical difference of these two choices will be analyzed below.}, 
\begin{multline}\label{eqn:CLFD_decomposition}
    \langle p' | T_{i}^{\alpha\beta}(0) | p\rangle = 2P^\alpha P^\beta A_i(-q^2) + \frac{1}{2}(q^\alpha q^\beta - q^2 g^{\alpha\beta}) D_i(-q^2) + 2M^2 g^{\alpha\beta}\bar{c}_i(-q^2) \\
    + \frac{M^4 \omega^\alpha \omega^\beta}{(\omega\cdot P)^2}S_{1i}(-q^2) + (V^\alpha V^\beta + q^\alpha q^\beta) S_{2i}(-q^2),
\end{multline}
where, $P = (p'+p)/2$, $q = p'-p$. Vector $V^\alpha$ is defined as $V^\alpha = \varepsilon^{\alpha \beta \rho \sigma}P_\beta q_\rho \omega_\sigma/(\omega\cdot P)$. This vector is perpendicular to $q^\alpha$, $P^\alpha$ as well as $\omega^\alpha$. 
Hermiticity excludes terms like $P^{\{\alpha} q^{\beta\}}$, $\omega^{\{\alpha} q^{\beta\}}$ or $\omega^{\{\alpha} V^{\beta\}}$. 
Form factors $S_{1i}$ and $S_{2i}$ are allowed by the reduced kinematic symmetries but forbidden by the full Poincaré symmetry. Therefore, they are called the spurious form factors. They are expected to vanish in the continuum limit. Note that the structure $V^{\alpha}V^{\beta}+q^\alpha q^\beta$ violates the current conservation. 

Using this decomposition Eq.~(\ref{eqn:CLFD_decomposition}), we can express the HMEs of the currents in terms of the GFFs as, 
\begin{align}
    A_i  
    =\,& \frac{1}{2(\omega\cdot P)^2} \omega_\alpha \omega_\beta t^{\alpha\beta}_i, \\
    D_i 
    =\,& \frac{2}{q^4} \big(q_{\alpha}q_{\beta}- V_\alpha V_\beta ) t^{\alpha\beta}_i, \\
    \bar c_i =\,& \frac{1}{4M^2 q^2}\Big\{
     (q_\alpha q_\beta - 3V_\alpha V_\beta + q^2 g_{\alpha\beta}) t^{\alpha\beta}_i - (M^2-\frac{1}{4}q^2) \frac{q^2}{(\omega\cdot P)^2}\omega_\alpha \omega_\beta t^{\alpha\beta}_i  \Big\}, \\
     S_1 =\,& \frac{1}{2M^4q^2}\Big[ (M^2-\frac{1}{4}q^2)(V_\alpha V_\beta + q_\alpha q_\beta - q^2 g_{\alpha\beta})t_i^{\alpha\beta} + 2q^2 P_\alpha P_\beta t_i^{\alpha\beta} - (M^2-\frac{1}{4}q^2)^2 \frac{q^2}{(\omega\cdot P)^2} \omega_\alpha \omega_\beta t_i^{\alpha\beta} \Big], \\
     S_2 =\,& \frac{1}{2q^4} \Big[ (q_\alpha q_\beta - q^2 g_{\alpha\beta} + 3V_{\alpha}V_\beta)t_i^{\alpha\beta} + (M^2-\frac{1}{4}q^2) \frac{q^2}{(\omega\cdot P)^2} \omega_\alpha\omega_\beta t_i^{\alpha\beta} \Big]. 
\end{align}
It is more transparent to write down the relations between the GFFs and the current components in standard light-front dynamics.  These relations are, 
\begin{align}
& t^{++}_i = 2(P^{+})^2 A_i(q^2_\perp), \label{eqn:t++_CLFD}\\
& t^{+a} = 2P^+P^a_\perp A_i(q^2_\perp), \label{eqn:t+a_CLFD}\\
& t^{ab}_i = 2 P^a_\perp P^b_\perp A_i(q^2_\perp) + \frac{1}{2}(q^a_\perp q^b_\perp - \delta^{ab}q^2_\perp) D_i(q^2_\perp) - 2M^2 \delta^{ab} \bar c_i(q^2_\perp) + \delta^{ab} q_\perp^2 S_{2i}(q_\perp^2), \label{eqn:tab_CLFD}\\
& t^{+-}_i = 2(M^2+P_\perp^2+\frac{1}{4}q_\perp^2)A_i(q^2_\perp) + q^2_\perp D_i(q^2_\perp) + 4M^2 \bar c_i(q_\perp^2), \label{eqn:t+-_CLFD} \\
& t^{--}_i = 2 \Big(\frac{M^2+P^2_\perp+\frac{1}{4}q_\perp^2}{P^+}\Big)^2 A_i(q_\perp^2) + 2 \Big(\frac{\vec q_\perp \cdot \vec P_\perp}{P^+}\Big)^2D_i(q^2_\perp) +  \frac{4M^4}{(P^+)^2}S_{1i}(q_\perp^2)  + \frac{4P^2_\perp q_\perp^2}{(P^+)^2} S_{2i}(q_\perp^2),  \label{eqn:t--_CLFD} \\
& t^{-a}_i = \frac{M^2+P^2_\perp+\frac{1}{4}q_\perp^2}{P^+}2P^a_\perp A_i(q_\perp^2) + \frac{\vec q_\perp \cdot \vec P_\perp}{P^+} q^a_\perp D_i(q^2_\perp) + \frac{2q^2_\perp }{P^+} P^a_\perp S_{2i}(q^2_\perp). \label{eqn:t-a_CLFD}
\end{align}
Here, $a = 1,2$ is the transverse index and $i=N, \pi$. 
Further simplification can be achieved by taking the Breit frame, $\vec P_\perp = \frac{1}{2}(\vec p_\perp + \vec p'_\perp) = 0$,
\begin{align}
& t^{++}_i= 2(P^{+})^2 A_i(q^2_\perp), \label{eqn:t++_CLFD_breit}\\
& t^{+a}_i = 0, \\
& t^{ab}_i = \frac{1}{2}(q^a_\perp q^b_\perp - \delta^{ab}q^2_\perp) D_i(q^2_\perp) - 2M^2\delta^{ab} \bar c_i(q^2_\perp) + \delta^{ab} q_\perp^2 S_{2i}(q_\perp^2), \\
& t^{+-}_i = 2(M^2+\frac{1}{4}q_\perp^2)A_i(q^2_\perp) + q^2_\perp D_i(q^2_\perp) + 4M^2\bar c_i(q^2_\perp), \\
& t^{--}_i = 2 \Big(\frac{M^2+\frac{1}{4}q_\perp^2}{P^+}\Big)^2 A_i(q_\perp^2) + \frac{4M^4}{(P^+)^2}S_{1i}(q_\perp^2), \\
& t^{-a}_i = 0, \label{eqn:t-a_CLFD_breit} \\
\Rightarrow \quad & \mathrm{tr}_\perp \tensor t_i = t^{11}_i + t^{22}_i = - \frac{1}{2} q^2_\perp D_i(q^2_\perp) - 4M^2 \bar c_i(q^2_\perp) + 2q_\perp^2S_{2i}(q_\perp^2), \\
& t^{12}_i = \frac{1}{2} q^1_\perp q^2_\perp  D_i(q^2_\perp).
\end{align}
Note that there are 5 independent HMEs and 5 GFFs. The physical GFFs $A_i, D_i, \bar c_i$ can only be extracted from combinations of the three ``good currents" $t^{++}_i$, $t^{+-}_i$ and $t^{12}_i$, without the contamination of the spurious form factors $S_{1,2}$. This observation is consistent with the perturbative results \cite{Cao:2023ohj}. 
For the total GFFs, in order for the Newton's third law $\sum_i \bar c_i(Q^2) = 0$ to hold, a non-trivial relation between these three good HMEs, 
\begin{equation}
\frac{M^2+\frac{1}{4}\vec q^2_\perp}{(P^+)^2} t^{++} + 2\frac{\vec q^2_\perp}{q^1_\perp q^2_\perp}t^{12} = t^{+-},
\end{equation}
has to be satisfied. We note that this relation is not automatically satisfied, except for $q^2_\perp = 0$. Therefore, Newton's third law (\ref{eqn:c-bar_sumrule}) holds when integrated over space on the light front $x^+=0$. For the scalar Yukawa model, $\bar c_i(Q^2)$ are constants in one-loop order. Therefore, Newton's third law (\ref{eqn:c-bar_sumrule}) is satisfied everywhere in this case. N.B. contrasting to the standard decomposition (\ref{eqn:HME_EMT}), violation of the current conservation in CLFD (\ref{eqn:CLFD_decomposition}) is caused by a divergent spurious GFF $S_{2}(Q^2) = \sum_i S_{2i}(Q^2)$, which is contained in $t^{11}$ and $t^{22}$. This analysis is consistent with the LCPT results in \ref{sect:LCPT}. 

Decomposition (\ref{eqn:CLFD_decomposition}) is not unique, and different forms are linearly related to each other with redefinitions
of the form factors. There is still arbitrariness in the Lorentz decomposition even if the physical part of the Lorentz structures is fixed \cite{Lorce:2015lna, Freese:2022yur, Cao:2023ohj}. For example, in our previous work \cite{Cao:2023ohj}, we adopted the Lorentz tensor $V^\alpha V^\beta$ instead of $V^\alpha V^\beta + q^\alpha q^\beta$ in Eq.~(\ref{eqn:CLFD_decomposition}). Yet another possible tensor is $P^{\{\alpha}\omega^{\beta\}}/(\omega \cdot P)$ \cite{Lorce:2015lna}, which can be expressed in terms of the present tensors as, 
\begin{equation}
\frac{P^{\{\alpha}\omega^{\beta\}}}{\omega \cdot P} = g^{\alpha\beta} + (M^2-\frac{1}{4}q^2)\frac{\omega^\alpha\omega^\beta}{(\omega\cdot P)^2} -  \frac{V^{\alpha}V^\beta + q^\alpha q^\beta}{q^2}. 
\end{equation}
Obviously, different decompositions are related to each other by redefinition of the spurious form factors $S_{1i}(-q^2)$ and $S_{2i}(-q^2)$ as well as the physical form factors $A_i(-q^2)$, $D_i(-q^2)$, $\bar c_i(-q^2)$. Note that the physical GFFs may also get redefined, which is in contrast to the vector current $J^\mu$ where the physical and spurious form factors do not mix. Since the spurious form factors in general carry uncanceled divergences, this redefinition also shuffles around the divergences.  

One of the pressing questions is how to choose a proper decomposition scheme such that the GFFs associated with the physical tensors, i.e. $P^\alpha P^\beta$,$q^\alpha q^\beta - q^2 g^{\alpha\beta}$ and $g^{\alpha\beta}$, correspond to the genuine physical GFFs. In a non-perturbative case, there is no easy answer to this problem, since even the physical form factors are only approximations of the true physical ones, unless a perturbative matching can be achieved. Fortunately, in the perturbative case, we can use the covariant results as a benchmark. It is evident that (\ref{eqn:CLFD_decomposition}) is the only decomposition scheme that is consistent with the LO perturbation theory shown in Sect.~\ref{sect:LCPT}, cf. Eqs.~(\ref{eqn:t++}--\ref{eqn:t+-}). The matching to the perturbative results also guarantees the exact cancellation of divergences in the physical GFFs. 
And, all divergences are confined within the spurious form factors $S_{1i}(-q^2)$ and $S_{2i}(-q^2)$. In particular, adoption of the spurious tensor $V^\alpha V^\beta + q^\alpha q^\beta$ ensures that the light-front energy density $T^{+-}$ is divergence free, which is consistent with our previous analysis using light-front energy conservation \cite{Cao:2023ohj}. 

Another interesting observation is that this tensor structure implies a hidden symmetry. The spurious terms within $t_i^{\mu\nu}$ depend only on $q^2_\perp$, 
\begin{align}
t^{+-}_\text{spur}(q^2_\perp)  =\,&  t^{12}_\text{spur}(q^2_\perp) = 0, \\
t^{11}_\text{spur}(q_\perp^2) =\,& t^{22}_\text{spur}(q_\perp^2) = q_\perp^2 S_{2i}(q^2_\perp),
\end{align}
i.e. they are isotropic in the transverse plane.When the theory is quantized on the light cone, a special direction, $z$ direction in standard light front coordinates, is introduced. This choice underlines the rise of the spurious terms. However, the rotational symmetry in the transverse plane is kept and there is no potential source, kinematical or dynamical, to break this symmetry. Therefore, the above observation is physically plausible. By contrast, the spurious tensor $V^\alpha V^\beta$ is anisotropic in the transverse direction. This symmetry suggests that our decomposition (\ref{eqn:CLFD_decomposition}) works beyond the leading-order perturbation theory. 

\section{Summary and outlooks}\label{sect:summary}

In this work, we compare the gravitational form factors of the scalar Yukawa theory obtained from covariant perturbation theory and from light-cone perturbation theory at the one-loop level. Our intention is to find a reliable way to extract the GFFs in light-front dynamics beyond the perturbative regime. In the light-front formalism, the full Poincaré symmetry is not manifest. As such, GFFs extracted from some current components of the EMT contain uncanceled divergences, which further lead to violation of conservation laws and known sum rules. 
Traditionally, the ``good currents" are used to extract physical form factors. In the case of the GFFs, the good currents $T^{++}$ (and $T^{+a}$) are not sufficient to obtain the $D$-term as well as the $\bar c$-term, i.e. the stress within hadrons. Using perturbation theory as a benchmark, 
we found that hadronic matrix elements $t^{++}_i$, $t^{+a}_i$, $t^{12}_i$ and $t^{+-}_i$ in light-cone perturbation theory reproduce those obtained from covariant perturbation theory whereas $t^{11}_i$, $t^{22}_i$ contains uncanceled divergences. Based on these analysis, we proposed a new covariant Lorentz decomposition of the energy-momentum tensor on the light front (\ref{eqn:CLFD_decomposition}), which takes into account the presence of a special direction $\omega^\mu$ when quantizing the theory. 

The new CLFD decomposition avoids the mixing of the uncanceled spurious divergence within the physical GFFs. In particular, the light-front energy density $T^{+-}$ does not contain spurious contributions, which is consistent with our previous derivation that this operator can be properly renormalized using the same set of counterterms that renormalize the light-front Hamiltonian \cite{Cao:2023ohj}. Furthermore, the new decomposition also introduces a new symmetry for the spurious terms on the transverse plane, viz. the breaking of the Poincaré symmetry is  isotropic in the transverse direction. 

One immediate application is the same system, the scalar Yukawa theory, in the non-perturbative regime. The present work overcomes the divergence issues encountered in the mass decomposition of the strongly coupled scalar nucleon \cite{Cao:2023ohj}. While the decomposition Eq.~(\ref{eqn:CLFD_decomposition}) differs from Eq.~(69) of our previous work, it can be shown that for the total GFF $D(Q^2)$ extracted from $T^{+-}$, these two decompositions give the same result.  

The method presented in this work is very general and can also be applied to other systems \cite{Brodsky:2000ii, More:2021stk, More:2023pcy, Rodini:2020pis, Metz:2021lqv, Freese:2022ibw, Freese:2022jlu, Eides:2023uox, Pasquini:2007xz, Amor-Quiroz:2023rke} and to hadrons with spin. One of the prime systems of interest is the nucleon. True nucleons are spin-1/2 hadrons. And they have one more set of GFFs $J_i$ (or $B_i$), representing the spin distribution inside the system \cite{Polyakov:2018zvc}. On the other hand, there are also spin-flip hadron matrix elements. Some of the spurious Lorentz structures contain zero-model contributions \cite{More:2023pcy}. Therefore, it is interesting to see whether covariant light-front dynamics can capture at least some of the zero-mode effects. 

\section*{Acknowledgements}

The authors acknowledge fruitful discussions with J. More, C. Mondal, S. Nair, K. Serafin, and G.~A.~Miller. 

 This work was supported in part by the National Natural Science Foundation of China (NSFC) under Grant No.~12375081, by the Chinese Academy of Sciences under Grant No.~YSBR-101, and by the US Department of Energy (DOE) under Grant No. DE-SC0023692. 

X. Z. is supported by new faculty startup funding by the Institute of Modern Physics, Chinese Academy of Sciences, by Key Research Program of Frontier Sciences, Chinese Academy of Sciences, Grant No.~ZDBS-LY-7020, by the Foundation for Key Talents of Gansu Province, by the Central Funds Guiding the Local Science and Technology Development of Gansu Province, Grant No.~22ZY1QA006, by Gansu International Collaboration and Talents Recruitment Base of Particle Physics (2023-2027), by International Partnership Program of the Chinese Academy of Sciences, Grant No.~016GJHZ2022103FN, by National Natural Science Foundation of China, Grant No.~12375143, by National Key R\&D Program of China, Grant No.~2023YFA1606903 and by the Strategic Priority Research Program of the Chinese Academy of Sciences, Grant No.~XDB34000000.
V.A.K. is supported by the Chinese Academy of Sciences President's International Fellowship Initiative Grant No. 2023VMA0010.

\end{document}